\documentclass[smallextended]{svjour3}      
\smartqed  % flush right qed marks, e.g. at end of proof
\usepackage{graphicx}
\usepackage{mathptmx}      
\usepackage{mathrsfs}
\usepackage[applemac]{inputenc}
\usepackage{latexsym}

\def\widebar{\overline}
\newcommand{\tr}{{\rm tr}}

\newcommand{\be}[1]{\begin{equation}\label{#1}}
\newcommand{\ee}{\end{equation}}
\newcommand{\ba}[1]{\begin{eqnarray}\label{#1}}
\newcommand{\ea}{\end{eqnarray}}

\journalname{General Relativity and Gravitation}

\begin{document}

\title{Self-acceleration and matter content in bicosmology from Noether Symmetries}

%\titlerunning{Short form of title}        % if too long for running head

\author{Mariam Bouhmadi-L\'opez \and
     Salvatore Capozziello  \and
        Prado Mart\'in-Moruno
}

%\authorrunning{Short form of author list} % if too long for running head

\institute{M. Bouhmadi-L\'opez \at
Department of Theoretical Physics, University of the Basque Country UPV/EHU, P.O. Box 644, 48080 Bilbao, Spain.
\\IKERBASQUE, Basque Foundation for Science, 48011 Bilbao, Spain.\\
              \email{mariam.bouhmadi@ehu.eus}       
              \and
              S. Capozziello \at
Dipartimento di Fisica ``E. Pancini",  Universit\`a di Napoli ``Federico II, Compl. Univ. di Monte Sant'Angelo, Via Cinthia 9, I-80126 Napoli, Italy.\\
Istituto Nazionale di Fisica Nucleare (INFN) Sez. di Napoli, Compl. Univ. di Monte Sant'Angelo, Via Cinthia 9, I-80126 Napoli, Italy.\\
Gran Sasso Science Institute, Viale F. Crispi 7, I-67100 L'Aquila, Italy.\\              
              \email{capozziello@na.infn.it}    
           \and
           P. Mart\'in-Moruno \at
              Departamento de F\'{\i}sica Te\'orica I, Ciudad Universitaria, Universidad Complutense de Madrid, E-28040 Madrid, Spain.\\
              \email{pradomm@ucm.es}
}

\date{Received: date / Accepted: date}
% The correct dates will be entered by the editor

\maketitle

\begin{abstract}
In bigravity, when taking into account  the potential existence of matter fields minimally coupled to the second gravitation sector, the dynamics of our Universe depends on some matter that cannot be observed in a direct way. In this paper, we assume the  existence of a Noether symmetry in bigravity cosmologies in order to constrain the dynamics of that matter. 
By imposing this assumption we obtain cosmological models with interesting phenomenology.
In fact, considering that our universe is filled with standard matter and radiation, we show that the existence of a Noether symmetry implies that 
either the dynamics of the second sector decouples, being the model equivalent to General Relativity, or the cosmological evolution of our universe tends to a de Sitter state with the vacuum energy in it given by the conserved quantity associated with the symmetry.
The physical consequences of the genuine bigravity models obtained are briefly discussed.
We also point out that the first model, which is equivalent to General Relativity, may be favored due to the potential appearance of instabilities 
in the second model.
\end{abstract}

\section{Introduction}

Bigravity theories, which are models of two mutually interacting dynamical metrics, 
were initially introduced by Isham, Salam, and Strathdee in the seventies \cite{Isham}.
These theories can be interpreted as describing
two universes interacting in a classical way through their gravitational effects
 \cite{Damour:2002ws}. Recently, they have attracted considerable attention since Hassan and Rosen
formulated a bigravity theory \cite{Hassan:2011zd} 
that is free of the Boulware--Deser ghost \cite{BD}. 
The formulation of this ghost-free bigravity theory was possible thanks 
to the development of a theory of massive gravity by de Rham, Gabadadze and Tolley that was  also potentially 
stable  \cite{deRham:2010ik,deRham:2010kj} 
(see references \cite{Hinterbichler:2011tt,deRham:2014zqa,Schmidt-May:2015vnx} for a reviews).
This massive gravity theory has 5 propagating modes, however,
some solutions may still present stability issues related to the scalar polarization 
\cite{Fasiello,Deser,Chamseddine:2013lid,Guarato:2013gba}. In addition,
there could be nontrivial gravitational effects in vacuum for massive gravity theories with a Friedmann-Lema\^itre-Robertson-Walker (FLRW)
background metric \cite{nonvacuum}.
Nevertheless, these potential
shortcomings would not necessarily affect the ghost-free bigravity theory \cite{nonvacuum,Fasiello:2013woa},
which could also be considered to be conceptually favoured against massive gravity
since it is a background independent theory. In bigravity the two gravitational sectors may be interpreted as weakly coupled worlds in the nomenclature introduced in reference \cite{Damour:2002ws}.
Unlike other modified gravity theories, the Solar System constraint on this type of theories are so tight that they automatically fulfils the constraint coming from the recently observed GW170817 and GRB 170817A events \cite{Baker:2017hug}.

Cosmological scenarios of the ghost-free bigravity theory have been studied assuming that there is only matter minimally coupled to one gravitational sector \cite{vonStrauss:2011mq,Volkov:2011an,Comelli:2011zm}.
These models can describe accelerated solutions in absence of a cosmological constant \cite{Akrami,Konnig:2013gxa}.
Nevertheless, as assuming the absence of matter coupled to one sector is a strong assumption \cite{Capozziello:2012re}, the possible presence of two sets of matter fields minimally coupled to each gravitational sector has also been considered \cite{Capozziello:2012re,Aoki:2013joa}.
The future cosmological evolution could approach a de Sitter state, a matter dominated universe, or even a spacetime future singularity when assuming ordinary matter in both sectors \cite{Aoki:2013joa}.
Moreover, perturbations in bimetric cosmologies for both cases have been investigated \cite{Comelli:2012db,Comelli:2014bqa,DeFelice:2014nja,Konnig:2014dna,Solomon:2014dua,Konnig:2014xva,Akrami:2015qga}.
It should be noted, however, that if the matter content minimally coupled to the second gravitational sector is completely arbitrary, then one could re-construct any desired cosmological evolution for our universe by simply fine-tuning that material content which is not directly observable \cite{Capozziello:2012re}. 
Therefore, it would be desirable to find an argument which would favor a particular kind of material content to be coupled to the second gravitational sector. 
On the other hand, more general couplings of matter to the gravitational sectors have also been considered \cite{Akrami:2013ffa,deRham:2014naa,Gumrukcuoglu:2015nua,Heisenberg:2015iqa}. Although some of these couplings could be free of the Boulware--Deser ghost below a cut-off scale \cite{deRham:2014naa}, we focus our attention on the ghost-free minimally coupled case.

Ghost-free multigravity theories \cite{Hinterbichler:2012cn,Luben:2016lku,Tamanini:2013xia},
general $f(R)$ kinetic terms for the metrics \cite{Nojiri:2012zu}, and Lanczos--Lovelock terms \cite{Hassan:2012rq} in higher dimensional generalizations  \cite{Do:2016uef} have also been investigated.
Nevertheless, in this paper,  we consider a theory of the form
\begin{eqnarray}\label{Sintro}
 S=S_{EH}(g)+\kappa\widebar S_{EH}(f)+m^2S_{int}(gf^{-1})
 +S_{m}(g,\phi_\alpha,\nabla_\alpha\phi_\alpha)
+\epsilon\widebar S_{m}(f,\widebar\phi_\alpha,\widebar\nabla_\alpha\widebar\phi_\alpha),
\end{eqnarray}
where we assume that the interaction term is independent of the derivatives of the metrics (and
it is of the particular form introduced in \cite{Hassan:2011zd}),
that $\phi_\alpha\neq\widebar\phi_\alpha$, $\kappa\neq0$, $m^2\neq0$ and $\epsilon$
are different constants. The interaction term leading to a ghost-free bigravity theory \cite{Hassan:2011zd} is the same as the Lagrangian terms giving mass to the graviton
in the ghost-free massive gravity theory \cite{deRham:2010ik,deRham:2010kj,Hassan:2011tf}. However, the characteristics and underlying philosophy of those theories is completely different. As it has been pointed out in reference \cite{Baccetti:2012bk}, if one insists in considering massive gravity as a particular limit of bimetric gravity, that limit would corresponds to setting $\kappa=\epsilon=0$. Such a limit must be taken carefully since there is not complete continuity in the parameter space of the theory \cite{Baccetti:2012bk}.

With these considerations in mind, in this paper we explore the possibility of fixing the role and dynamics of the minimally coupled matter in bigravity. 
The existence of symmetries has been a guidance principle in the construction of different models in physics. 
Usually one assumes a particular symmetry which is physically well motivated to restrict the forms of the possible  Lagrangian models of the theory. 
The  Noether Symmetry Approach consists in giving a class of Lagrangians,  restricting to those presenting Noether symmetries, and integrating the related dynamics \cite{Noether}. Specifically, in the context of  theories of gravity \cite{extended1,extended2}
the Noether Symmetry Approach has been proven to be a useful method for obtaining exact
solutions in cosmological scenarios \cite{Noether1,Noether2}.
The method consists in assuming the existence of an arbitrary Noether symmetry for a family of Lagrangians in cosmological scenarios. Since a symmetry does not always exist, the method select both the form of the Lagrangians that have a symmetry (fixing, in particular, the form of couplings and potentials) and the particular symmetry (i.e. the associated conserved quantity). 
Moreover,  once the symmetry is known, the equations of motion can be easily integrated. 
The Noether Symmetry Approach has allowed to select particular theories of interest and to obtain exact cosmological solutions in several cases (see for example \cite{felicia,basilakos,defelice,nesseris}). It is worth noticing that symmetries have always a physical meaning and their existence  discriminates between physical and unphysical models besides allowing integrability of related dynamical systems. 
{In the present case, the existence of symmetries determines the matter content of the bigravity system. 
In particular, the existence of a Noether symmetry implies that either the dynamics of the second cosmological sector decouples, 
being the model equivalent to general relativity, or  the cosmological evolution of the  observed universe evolves towards a de Sitter 
state with the vacuum energy (the cosmological constant) determined by the conserved quantity associated to the symmetry. 
In other words, the matter content (and the source) of  the cosmological dynamical model is determined by the symmetry.
On the contrary, the absence of symmetry does not allow any conclusion on matter content and cosmological constant.
Another important example is discussed in \cite{Noether2} where minisuperspace models, related to quantum cosmology,  are  taken into account. There, the  presence of Noether symmetries determines oscillating behaviors in the wave function of the universe, solution of the Wheeler-de Witt equation. In this case, according to the Hartle criterion, correlated quantities can be identified and conserved momenta imply that observable universes are selected. If the symmetries do not exist, it is not possible to state if observable models can be found or not. In this sense, symmetries are a sort of criterion to  select physical models}. 

In bigravity the situation is different than in the case of extended theories of gravity. Here, the Lagrangian is already fixed and belongs to the family of ghost-free Lagrangians, which  includes only  four free parameters, but the kind of matter fields, minimally coupled to the second gravitational sector,  is completely arbitrary, entailing a great freedom \cite{Capozziello:2012re}. 
Hence, we consider that the material content, minimally coupled to the second sector, should be such that there is a Noether symmetry for the resulting cosmological model.
Therefore, in this paper, we will apply the Noether Symmetry Approach to bigravity cosmology not only to select the material content minimally coupled to the second gravitational sector
(when assuming the presence of a suitable amount of ordinary matter minimally coupled to our sector) but also to fix some of the free parameters appearing in the interaction Lagrangian. 

Specifically, searching for Noether symmetries in bigravity cosmology has some advantages that can be listed as follows: $i)$ it allows to restrict the range of free parameters according to the criterion of existence or not of symmetry. This seems an arbitrary  assumption,  but it is motivated by the fact that the presence of  conserved quantities allows the reduction of dynamical systems, and, in principle,  their   exact integration,  if the number of first integrals  is the same as the number of  configuration space variables; $ii)$ if  the observed universe is filled with standard matter and radiation, the presence of Noether symmetries allows the decoupling with respect to the second gravitational sector, being the model equivalent to general relativity. This decoupling is not guaranteed,   if a symmetry is not present: in fact, the  decoupling strictly relies on the possibility to integrate the dynamical system; $iii)$ finally, if the observed universe evolves towards a de Sitter phase, the conserved quantity related to the Noether symmetry allows to fix the vacuum energy (i.e. the cosmological constant) of the model.

{Despite the advantages of the existence of a Noether symmetry that we have just discussed, we have to comment also 
a potential difficulty of some bigravity models that we obtain in this paper. We start by considering a particular bigravity Lagrangian
because it propagates the correct number of modes and is, therefore, free of the  Boulware--Deser ghost. However, we obtain that the purely bigravity
models, which are not equivalent to General Relativity (GR) due to the decoupling of the gravitational sectors, with a Noether symmetry, present gravitational couplings
in both sectors with different signs. So these models could still lead to other instabilities. This conclusion would point out that GR is the only stable bigravity model with a Noether symmetry. This statement needs further detailed analysis.}

The layout of the paper is  as follows: In section \ref{bc}, we present the basic formalism of cosmological solutions in bigravity.
We summarize how to obtain the Friedmann equations in subsection~\ref{bc1} and we construct the point-like Lagrangian in subsection~\ref{bc2}.
In section \ref{NS}, we consider the Noether  Symmetry Approach in the context of cosmological bigravity solutions in order to fix the material content minimally coupled to the second gravitational sector. We review the Noether  Symmetry Approach in subsection \ref{s31}, apply it to bigravity cosmological models in subsection \ref{s32}, and discuss the particular solutions that we obtain in subsection \ref{s33}.
In section \ref{gravitation}, we discuss the potential issues of some particular cosmological solutions with a Noether symmetry. We summarize the results in section \ref{summary} and relegate some considerations about the definition of the point-like Lagrangian to appendix \ref{symmetric}.

%%%%%%%%%%%%%%%%%%%%%%%%%%%%%%%%%
%%%%%%%%%%%%%%%%%%%%%%%%%%%%%%%%%
\section{Bicosmology}\label{bc}
In this section, we include some  results about cosmological scenarios
of the ghost-free bimetric theory of gravity \cite{Hassan:2011zd}.
In the first place, in the subsection~\ref{bc1}, 
we summarize the equations of motion retrieved from the action of the theory and, in particular, the modified Friedmann equations, obtained by substituting the cosmological ansatz in the modified Einstein equations. 
In  subsection~\ref{bc2}, we re-obtain a point-like Lagrangian
by considering the cosmological ansatz in the general action of the theory. As we will show, the modified
Friedmann equations can also be extracted by varying the point-like Lagrangian, therefore,
being both procedures compatible. Finally, in subsection~\ref{bc3}, we present an equivalent but non-degenerate point-like Lagrangian that will be suitable for the subsequent analysis.

%%%%%%%%%%%%%%%%%%%%%%%%%%%%%%%%%
%%%%%%%%%%%%%%%%%%%%%%%%%%%%%%%%%
\subsection{Bimetric cosmology}\label{bc1}
The interaction Lagrangian of the ghost-free bigravity theory formulated by Hassan and Rosen is a 
function of the matrix $\gamma$, implicitely defined as
\begin{equation}
 \gamma^\mu{}_\rho\gamma^\rho{}_\nu =g^{\mu\rho}f_{\mu\rho}.
\end{equation}
The action reads \cite{Hassan:2011zd}
\begin{eqnarray}\label{action}
S&=&\frac{M_P^2}{2}\int d^4x\sqrt{-g} R(g) +\frac{\kappa M_P^2}{2}\int d^4x\sqrt{-f} R (f)   
-m^2\int d^4x\sqrt{-g} \sum_{n=0}^{4}\beta_n\,e_n(\gamma)
\nonumber\\
&+&
\int d^4x\sqrt{-g}  \,L_\mathrm{m}\left(\phi_\alpha,\,\nabla_\beta\phi_\alpha\right)
+\int d^4x\sqrt{-f} \,\widebar L_\mathrm{m}\left(\widebar\phi_\alpha,\,\widebar\nabla_\beta\widebar\phi_\alpha\right),
\end{eqnarray}
where the parameter $\epsilon$ appearing in equation (\ref{Sintro}) has been absorved in $\widebar L_\mathrm{m}$.
On the above expression, the elementary symmetric polynomials are \cite{Hassan:2011vm}
\begin{eqnarray}\label{symm}
e_0(\gamma) &=&1,\\
e_1(\gamma) &=&\tr[\gamma],\\
e_2(\gamma) &=&\frac{1}{2}\left(\tr[\gamma]^2-\tr[\gamma^2]\right),\\
e_3(\gamma) &=&\frac{1}{6}\left(\tr[\gamma]^3-3\tr[\gamma]\tr[\gamma^2]+2\tr[\gamma^3]\right),\\
e_4(\gamma) &=&{\rm det}(\gamma),
\end{eqnarray}
and the fields $\phi_\alpha$ and $\widebar\phi_\alpha$ are minimally coupled to $g$ and $f$, respectively. In addition, $M_P$ is the Planck mass\footnote{It must be noted that $M_P$ is related with the gravitational coupling appearing in the modified Einstein equations (\ref{Einstein1}). However, it could be considered that the physical Planck mass is the one of the massless spin-2 mode mediating the gravitational force, that is $M_P (1 + \kappa)$.}, $\kappa$ a dimensionless constant, and $m$ and $\beta_n$ are (free) constants of the model with dimensions of mass and inverse mass squared, respectively.
It must be noted that although we have not explicitly written a cosmological constant for each kinetic term
in the action (\ref{action}), those cosmological constants have been absorbed in the
$e_0$ and $e_4$ terms of the interaction Lagrangian since they are equivalent to cosmological constant contributions for the $g$-space
and $f$-space \cite{Hassan:2011vm,Baccetti:2012re}. 
The variation of the action (\ref{action}) with respect to the two metrics leads to two
sets of modified Einstein equations. These are~\cite{vonStrauss:2011mq,Baccetti:2012bk}
\begin{equation}\label{Einstein1}
 G^{\mu}{}_{\nu} =\frac{1}{M_P^2} \left(T^{({\rm m})\mu}{}_{\nu}+T^{\mu}{}_{\nu}\right),
\end{equation}
and
\begin{equation}\label{Einstein2}
\widebar{G}^{\mu}{}_{\nu}  =
\frac{1}{\kappa M_P^2}\left(\widebar T^{({\rm m})\mu}{}_{\nu} + \widebar{T}^{\mu}{}_{\nu}\right),
\end{equation}
where
\begin{eqnarray}\label{Tg}
 T^{\mu}{}_{\nu}&=& -m^2\left[\beta_0+\beta_1\,e_1(\gamma)+\beta_2\,e_2(\gamma)+\beta_3\,e_3(\gamma)\right]\; \delta^\mu{}_\nu\nonumber\\
 &+ &m^2\left[\beta_1+\beta_2\,e_1(\gamma)+\beta_3\,e_2(\gamma)\right] \gamma^\mu{}_\nu \nonumber\\
&-&
m^2\left[\beta_2+\beta_3\,e_1(\gamma)\right] \{\gamma^2\}^\mu{}_\nu +
\beta_3 \,\{\gamma^3\}^\mu{}_\nu,
\end{eqnarray}
and
\begin{eqnarray}\label{Tf}
 \widebar{T}^{\mu}{}_{\nu}&=&-m^2\sqrt{gf^{-1}}\left\{\left[\beta_1+\beta_2\,e_1(\gamma)+\beta_3\,e_2(\gamma)\right] \gamma^\mu{}_\nu \right. \nonumber\\
 &-&\left.\left[\beta_2+\beta_3\,e_1(\gamma)\right] \{\gamma^2\}^\mu{}_\nu + \beta_3\, \{\gamma^3\}^\mu{}_\nu \right\}
  -m^2\beta_4.
\end{eqnarray}
The indexes of equations (\ref{Einstein1}) and (\ref{Einstein2}) must be raised and lowered using $g$ and $f$, 
respectively. In addition, $T^{({\rm m})\mu}{}_{\nu}$ and $\widebar T^{({\rm m})\mu}{}_{\nu}$ correspond to the matter energy-momentum tensors of the $g$-space and $f$-space, respectively,

Now, let us consider a cosmological scenario. We assume that the metrics can be written as follows:
\begin{equation}\label{metric-g}
ds_g^2 = - N(t)^2 dt^2 + a(t)^2 \left[ {dr^2\over 1- k r^2} + r^2 d\Omega_{(2)}^2 \right],
\end{equation}
and
\begin{equation}\label{metric-f}
ds_f^2 = - \widebar N(t)^2 \; dt^2 + b(t)^2 \left[ {dr^2\over 1- k r^2} + r^2 d\Omega_{(2)}^2 \right],
\end{equation}
where $d\Omega_{(2)}^2=d\theta^2 + \sin^2\theta\; d\phi^2$  and $k=0,\pm 1$ is the spatial curvature parameter. More general cosmological ansatzs have been also considered in the literature (cf. references \cite{Nersisyan:2015oha} and \cite{Garcia-Garcia:2016dcw}).
The modified Friedman equations for each space can be obtained
substituting the ansatzs (\ref{metric-g}) and (\ref{metric-f}) in equations (\ref{Einstein1}) and (\ref{Einstein2}).
That can be done either by brute force \cite{vonStrauss:2011mq} or 
noting that (\ref{metric-g}) and (\ref{metric-f}) are related through the generalized
Gordon ansatz \cite{Baccetti:2012ge}. Following any of the two procedures, one obtains
\begin{equation}\label{poli}
 e_1=3\frac{b}{a}+\frac{\widebar N}{N},\,\, e_2=3\frac{b^2}{a^2}+3\frac{\widebar N}{N}\frac{b}{a},\,\, 
e_3=\frac{b^3}{a^3}+3\frac{\widebar N}{N}\frac{b^2}{a^2}.
\end{equation}
The modified Friedmann equations are
\begin{equation}\label{Fried-g}
H_\mathrm{g}^2+\frac{k}{a^2}=\frac{1}{3\,M_P^2}\left[\rho_{\rm bg}(b/a)+\rho_\mathrm{m}(a)\right],
\end{equation}
and
\begin{equation}\label{Fried-f}
H_\mathrm{f}^2+\frac{k}{b^2}=\frac{1}{3\,\kappa M_P^2}\left[\widebar\rho_{\rm bg}(a/b)+\widebar\rho_\mathrm{m}(b)\right],
\end{equation}
where the effective energy density due to the interaction in each space has been defined as \cite{Baccetti:2012ge}
\begin{equation}\label{rho-g}
 \rho_{\rm bg}=m^2\left(\beta_0+3\beta_1\frac{b}{a}+3\beta_2\frac{b^2}{a^2}+\beta_3\frac{b^3}{a^3}\right),
\end{equation}
\begin{equation}\label{rho-f}
 \widebar \rho_{\rm bg}=m^2\left(\beta_4+3\beta_3\frac{a}{b}+3\beta_2\frac{a^2}{b^2}+\beta_1\frac{a^3}{b^3}\right),
\end{equation}
and the Hubble functions are
$H_\mathrm{g}=\dot a/(N\,a)$ and $H_\mathrm{f}=\dot b/(\widebar N\,b)$, respectively, with
$\dot{} \equiv d/dt$. 

On the other hand, due to the diffeomorphism invariance of the matter actions appearing in the
action~(\ref{action}), the stress-energy tensors of the matter components of both spaces are conserved. 
Therefore one has
$\dot\rho_\mathrm{m}/N+3H_\mathrm{g}[1+w_\mathrm{m}(a)]\,\rho_\mathrm{m}=0$ and 
$\dot{\widebar\rho}_\mathrm{m}/\widebar N+3H_\mathrm{f}[1+\widebar w_\mathrm{m}(b)]\,\widebar\rho_\mathrm{m}=0$.
Moreover, taking into account the Bianchi identities together with these conservations, 
the effective stress energy tensors (\ref{Tg}) and (\ref{Tf}) must be also
conserved. 
This leads to the Bianchi-inspired constraint\footnote{The use of the generalized Gordon ansatz implies that we only consider the branch of the Bianchi-inspired constraint given by equation (\ref{Bi}). However, it should be noted that there is a second branch that leads to solutions whose scale factors are  proportional to $a(t)$ and $b(t)$ (with fixed constant of proportionality) \cite{vonStrauss:2011mq}. As pointed out in reference \cite{nonvacuum}, restriction to the second branch implies the elimination of all nontrivial interaction terms, i.e., those which are not equivalent to a cosmological constant. Moreover, cosmological solutions of this branch have a nonlinear instability \cite{DeFelice:2012mx}.} \cite{Volkov:2011an,vonStrauss:2011mq,Baccetti:2012ge}
\begin{equation}\label{Bi}
\frac{\dot b(t)}{\widebar N(t)}=\frac{\dot a(t)}{N(t)}.
\end{equation}
As is already well known, this implies that the Hubble function of the $f$-space can be expressed as 
$H_\mathrm{f}=\dot a/(N\,b)$. Therefore, the Friedmann equations of both spaces, equations (\ref{Fried-g}) and 
(\ref{Fried-f}), are coupled \cite{Volkov:2011an,vonStrauss:2011mq}. This allows us to write
the Friedmann equation of the second universe (\ref{Fried-f}) as
\begin{equation}\label{Fried-f2}
\frac{a^2}{b^2}\left(H_\mathrm{g}^2+\frac{k}{a^2}\right)=\frac{1}{3\,\kappa M_P^2}\left[\widebar\rho_{\rm bg}(a/b)+\widebar\rho_\mathrm{m}(b)\right],
\end{equation}
which can be combined with equation (\ref{Fried-g}) to remove the term $H_\mathrm{g}^2$. 
Considering a general material content
in both universes, this procedure leads to \cite{Capozziello:2012re}
\begin{equation}\label{polinomio}
c_4\Gamma^4+c_3 \Gamma^3-\widebar D\,\widebar\rho_\mathrm{m}(b) \Gamma^3  +c_2 \Gamma^2
+D\rho_\mathrm{m}(a) \Gamma
+c_1 \Gamma-c_0 =0,
\end{equation}
where $\Gamma=b/a$, $c_4=\beta_3/3$, $c_3=\beta_2-\beta_4/(3\kappa)$, 
$c_2=\beta_1-\beta_3/\kappa$, $c_1=\beta_0/3-\beta_2/\kappa$, 
$c_0=\beta_1/(3\kappa)$, $D=1/(3m^2)$,
$\widebar D=1/(3m^2\kappa)$. This equation can be seen as an algebraic equation in $b$ and, therefore, it can be solved to obtain $b(a)$ at least in principle.
Solutions for the case in which the second universe is empty are generically easier to obtain, as one can define 
the quantity $\Gamma=b/a$ as in reference \cite{vonStrauss:2011mq} and then equation (\ref{polinomio}) 
becomes a quartic equation in $\Gamma$  (it could even be simpler for models in which some parameters $\beta_i$ vanish \cite{vonStrauss:2011mq}).
Once one obtains $b(a)$ from equation (\ref{polinomio}),
it can be substituted in the Friedmann equation to study the dynamics of the universe by integrating $a(t)$.

%%%%%%%%%%%%%%%%%%%%%%%%%%%%%%%%%
%%%%%%%%%%%%%%%%%%%%%%%%%%%%%%%%%
\subsection{The point-like Lagrangian}\label{bc2}

In the previous subsection, we have written the general equations for the dynamics of the metrics, that
is the modified Einstein equations. Then we have restricted our attention to solutions described by two FLRW geometries. 
Nevertheless, one could have also studied the problem considering a point-like 
Lagrangian, which leads to dynamics in a minisuperspace.
To obtain such a point-like Lagrangian, one has to substitute the cosmological metrics given by equations~(\ref{metric-g})
and (\ref{metric-f}) in action~(\ref{action}). 
Therefore, by replacing the expressions (\ref{poli}) in the interaction term of the Lagrangian (\ref{action}) and
integrating by parts the Einstein-Hilbert Lagrangian of action~(\ref{action}), one  gets the following
point-like Lagrangian:
\begin{eqnarray}\label{pl1}
{\cal L}&=& N a^3\left[-3M_P^2\frac{\dot a^2}{N^2a^2}+3M_P^2\frac{k}{a^2}-\rho_{\rm bg}(b/a)-\rho_{\rm m}(a)\right]\nonumber\\
&+& \widebar N b^3\left[-3\kappa M_P^2\frac{\dot b^2}{\widebar N^2 b^2}+3\kappa M_P^2\frac{k}{b^2}-\widebar\rho_{\rm bg}(a/b)-
\widebar\rho_{\rm m}(b)\right],
\end{eqnarray}
with $\rho_{\rm bg}(b/a)$ and $\widebar\rho_{\rm bg}(a/b)$ have been defined in equations (\ref{rho-g}) and (\ref{rho-f}), 
respectively.
This is the same Lagrangian as that presented in reference \cite{Fasiello:2013woa}, but we have
split the interaction term in $\rho_{\rm bg}(b/a)$ and $\widebar\rho_{\rm bg}(a/b)$ for convenience.
It must be noted that the Lagrangian (\ref{pl1}) is defined in the tangent space,
 ${\cal TQ}\equiv\{N,\,\dot N,\,\widebar N,\,\dot{\widebar N},\,a,\,\dot a,\,b,\,\dot b\}$, coming from the configuration space
${\cal Q}\equiv\{N,\,\widebar N,\,a,\,b\}$.

Varying the Lagrangian (\ref{pl1}) with respect to $N$ and $\widebar N$, we obtain the modified Friedmann
equations which  are equations~(\ref{Fried-g}) and (\ref{Fried-f}), respectively. Now, the variation
with respect to the scale factor $a$ leads to
\begin{eqnarray}\label{aceleration}
-\frac{1}{M_P^2}\left(\rho_{\rm bg}+\frac{a}{3}\rho'_{\rm bg}+
\frac{1}{3}\frac{b^3}{a^2}\frac{\widebar N}{N}\widebar \rho'_{\rm bg}+\rho_{\rm m}+\frac{a}{3}\rho_{\rm m}'\right)
=
 -2\frac{1}{a}\frac{d^2a}{dt_{ca}^2}-\left[\frac{1}{a}\frac{da}{dt_{ca}}\right]^2-\frac{k}{a^2},
\end{eqnarray}
with $dt_{ca}\equiv N\,dt$ and $'\equiv\partial/\partial a$. The conservation equation can be written as
$\rho_{\rm m}+\frac{a}{3}\rho'_{\rm m}=-p_{\rm m}$. Therefore, if the interaction term vanishes (for $m^2=0$),
the acceleration equation (\ref{aceleration}) will be equivalent to considering the derivative of Friedmann equation 
and the conservation of
the fluid, as in usual   GR; i.e. the standard Raychaudhuri equation. By similarity to what happens in GR, we can define an effective pressure, $p_{\rm bg}$, as follows
\begin{equation}\label{pbg}
 p_{\rm bg}=-\left(\rho_{\rm bg}+\frac{a}{3}\rho'_{\rm bg}+
\frac{1}{3}\frac{b^3}{a^2}\frac{\widebar N}{N}\widebar \rho'_{\rm bg}\right).
\end{equation}
Combining the derivative of the Friedmann equation~(\ref{Fried-g}) with equation~(\ref{aceleration}),  one obtains
\begin{equation}\label{Bii}
\frac{N}{\widebar N}=\frac{\dot a}{\dot b},
\end{equation}
which is exactly the Bianchi-inspired constraint (\ref{Bi}).
Therefore, this procedure is equivalent to that presented in the previous subsection.

Nevertheless, it can be noted that the determinant of the Hessian\footnote{The Hessian of the Lagrangian is defined as $H_{ij}=\partial^2{\cal L}/{(\partial \dot q^i}{\partial \dot q^j)}$.}  of the  Lagrangian~(\ref{pl1}) vanishes and, therefore, the Lagrangian is degenerate \cite{Noether}
in the tangent space ${\cal TQ}\equiv\{N,\,\dot N,\,\widebar N,\,\dot{\widebar N},\,a,\,\dot a,\,b,\,\dot b\}$.
This fact can be problematic when applying certain procedures, as the Noether  Symmetry Approach or 
the quantization of the corresponding Hamiltonian. Therefore, one can interpret the fact that the Bianchi inspired
constraint (\ref{Bi}) is obtained by deriving the equations of motion and requiring their compatibility,
as being a consequence of the large number of equations of motion provided by an initial degenerate Lagrangian.
The information is there, but it is redundant.
As we will see in the next section, one can define a non-degenerate Lagrangian containing the information encoded in equation (\ref{Bi}) from the beginning.

%%%%%%%%%%%%%%%%%%%%%%%%%%%%%%%%%
%%%%%%%%%%%%%%%%%%%%%%%%%%%%%%%%%
\subsection{Non-degenerate point-like Lagrangian}\label{bc3}
The Lagrangian~(\ref{pl1}) is degenerate because it is independent of $\dot N$ and $\dot{ \widebar N}$.
Therefore, to find a non-degenerate point-like Lagrangian we need to remove the dependence on the non-dynamical variables $N$ and $\widebar N$ in ${\cal L}$. 
In the first place, we note that we can use the temporal gauge of freedom to fix $N=1$ without loss of generality; then the cosmic time of the $g$-universe will be $t$. 
As in this theory we have only one global invariance under changes of coordinates, this choice already fixes the temporal gauge freedom and, therefore, we cannot
freely choose $\widebar N$. 
Indeed, we already know that $\widebar N$ has to be given by equation (\ref{Bi}). Therefore, we take
\begin{equation}\label{N}
N=1,\qquad\widebar N=\frac{\dot b}{\dot a},
\end{equation}
in the Lagrangian (\ref{pl1}) to obtain\footnote{Although, as we have discussed in the previous section, the information of the Bianchi-inspired constraint was already included in the Lagrangian (\ref{pl1}), by considering the equations (\ref{N}), we remove redundant information and with it the degeneracy of the Lagrangian.}
\begin{eqnarray}\label{pl2}
 {\cal L}&=&- \left[3M_P^2\,a\,\dot a^2-3M_P^2k\,a+a^3\rho_{\rm bg}(b/a)+a^3\rho_{\rm m}(a)\right]\nonumber\\
&-& \left[3\kappa M_P^2\,b\,\dot a\, \dot b-3\kappa M_P^2k\,b\frac{\dot{b}}{\dot{a}}+b^3\frac{\dot b}{\dot a}\widebar\rho_{\rm bg}(a/b)+b^3\frac{\dot b}{\dot a}
\widebar\rho_{\rm m}(b)\right] .
\end{eqnarray}
It has to be emphasized that this Lagrangian is not symmetric under inter-change of the $g$-universe and $f$-universe, as it has been the case until fixing the gauge of freedom. The reason is that the gauge fixing (\ref{N}) breaks the symmetry between both gravitational sectors, so ${\cal L}$ is not invariant under the inter-change $b\leftrightarrow a$.
In particular, we are choosing to express the physics
in terms of our cosmic time. As it is suggested in the appendix \ref{symmetric}, a symmetry breaking gauge fixing seems to be necessary to obtain
a non-degenerate Lagrangian. 

In order to check that the Lagrangian (\ref{pl2}) describes the same physical situation as (\ref{pl1}), one
should obtain the same information; i.e. dynamical evolution, as in the previous section.
In fact, varying the Lagrangian with respect to $b$ and after some simplifications, one obtains
\begin{equation}\label{Ff}
\left(\frac{\dot a}{b}\right)^2+\frac{k}{b^2}=\frac{1}{3\kappa M_P^2}\left[\widebar\rho_{\rm bg}(a/b)+\widebar\rho_{\rm m}(b)\right].
\end{equation}
This is just the Friedmann equation of the second universe, equation (\ref{Fried-f}), with $H_f$ written in terms of $\dot a$ through
equation~(\ref{N}). Moreover, varying with respect to $a$, simplifying, substituting
equation (\ref{Ff}) when needed, and simplifying again, one  gets
\begin{equation}\label{Fg}
 2\frac{\ddot a}{a}+\left(\frac{\dot a}{a}\right)^2+\frac{k}{a^2}=-\frac{1}{M_P^2}\left[p_{\rm m}(a)+p_{\rm bg}(b/a)\right],
\end{equation}
with $p_{\rm bg}$ given by equation (\ref{pbg}), which is equation~(\ref{aceleration}). Thus, the system of equations
is equivalent to that obtained from the Lagrangian (\ref{pl1}).

On the other hand, as the non-degenerate point-like Lagrangian has no explicit dependence on $t$, the energy is conserved. That is
\begin{eqnarray}\label{EL}
 E_{\cal L}&=&\frac{\partial  {\cal L}}{\partial \dot a}\dot a+ \frac{\partial  {\cal L}}{\partial \dot b}\dot b- {\cal L}\nonumber\\
 &=&-3M_P^2a^3\left[\frac{\dot a^2}{a^2}+\frac{k}{a^2}-\frac{\rho_{\rm bg}+\rho_{\rm m}}{3M_P^2}\right]
-3\,\kappa M_P^2b^3\frac{\dot b}{\dot a}\left[\frac{\dot a^2}{b^2}+\frac{k}{b^2}-\frac{\widebar\rho_{\rm bg}+\widebar\rho_{\rm m}}{3\,\kappa M_P^2}\right]\nonumber\\
&=&0,
\end{eqnarray}
where we know that this function vanishes for compatibility with the Friedmann equation. 
Moreover, the Hamiltonian function of the system can be obtained from the energy (\ref{EL}) when re-expressing the ``velocities" in terms of the momenta; therefore, it has to vanish due to the diffeomorphism invariance.
Furthermore, although equation (\ref{EL})  could suggest that the energy of each universe
is conserved separately, this is not the case because writing
$E_{\cal L}= E_{{\cal L}_1}+ E_{{\cal L}_2}$, $E_{{\cal L}_1}$ cannot be obtained using only ${\cal L}_1$
(it has contributions from ${\cal L}_2$), and vice versa. Anyway, it is important to emphasize that
the consideration of the equation implied by $ E_{\cal L}=0$ is equivalent to one of the equations of motion of the system,
and it would be easier to calculate in general.

%%%%%%%%%%%%%%%%%%%%%%%%%%%%%%%%%
%%%%%%%%%%%%%%%%%%%%%%%%%%%%%%%%%
\section{Bigravity cosmologies with a Noether symmetry}\label{NS}
As we have discussed in the introduction, the dynamics of our Universe in bigravity is determined not only by the material content minimally coupled to our gravitational sector but also by the material content minimally coupled to the other sector. As that set of matter fields cannot be measured in a direct way, it seems that any cosmological dynamics could be described by fine-tuning it. Therefore, the theory would loss some predictive power if the nature of the  ``hidden" matter content cannot be fixed by fundamental principles.
We assume that such a fundamental principle is the existence of an additional Noether symmetry for cosmological solutions. 
We consider that this assumption is more robust than assuming an empty second gravitational sector from the beginning. The existence of symmetries commonly simplifies the treatment of physical problems. As we will show, however, our universe cannot have a Noether symmetry during its whole evolution, but it can tend to a state where this symmetry is present and the kind of material content is, therefore, fixed. The phenomenological interest of the solutions obtained are, at the end of the day, the stronger indication in favor of the considered assumption.

In this section, we review the Noether  Symmetry Approach formalism in subsection \ref{s31}, we apply this formalism to general bigravity cosmological solutions in subsection \ref{s32}, and, we particularize the study considering a specific kind of material content minimally coupled to our sector in subsection \ref{s33} extracting the particular solutions.

%%%%%%%%%%%%%%%%%%%%%%%%%%%%%%%%%
%%%%%%%%%%%%%%%%%%%%%%%%%%%%%%%%%
\subsection{The Noether Symmetry Approach}\label{s31}
The motivation behind the use of the Noether  Symmetry Approach in cosmology is to select a particular modified gravity model from the general class of theories considered
and to find exact cosmological solutions for those model. From a classical point of view, the selected model may be interpreted as being favored against other Lagrangians since it has an additional symmetry \cite{Noether}. This interpretation entails, of course, the underlying assumption that the most suitable theory of the family of theories  is that implying the existence of more symmetries.
On the other hand, from a quantum point of view, the presence of Noether symmetries allows to select observable universes by the Hartle criterion, since the absence of Noether symmetries gives rise to non-correlated variables and then to non-observable universes  \cite{Noether2} .

In order to apply the Noether  Symmetry Approach,  one needs a non-degenerate point-like
Lagrangian ${\cal L}(q_i,\,\dot q_i) $ which is independent of time.
Then, one supposes the existence of  a Noether symmetry, which implies that \cite{Noether}
\begin{equation}\label{Lie0}
 L_X{\cal L}(q_i,\,\dot q_i)=0,
\end{equation}
where $X$ is a vector field defined on the tangent space ${\cal TQ}\equiv \{q^i,\,\dot{q}^i\}$, that is 
\begin{equation}\label{X0}
X=\xi^i(q^i)\frac{\partial}{\partial q^i}+\dot{\xi}^i(q^i)\frac{\partial}{\partial \dot{q}^i}\,,
\end{equation}
and $L_X$ is the Lie derivative along the direction $X$. 
Equation (\ref{Lie0}) gives rise to a system of partial differential equations whose solution is not unique. The  solutions of equation (\ref{Lie0})  fix the vector components
$\xi^i(q^i)$ (and consequently $\dot{\xi}^i(q^i)$ and the symmetry)  and the functional form of   the Lagrangian  ${\cal L}(q_i,\,\dot q_i)$, that is the couplings and potentials  (see \cite{Noether} for details).
Besides, the Lagrangian, and then the dynamics,  are associated to a conserved quantity $\Sigma_0$  that can be used
to integrate the equations of motion, being
\begin{equation}
\frac{d}{dt}\left(\xi^i\frac{\partial {\cal L}}{\partial {\dot q}^i}\right)=L_X{\cal L}\,,
\end{equation}
and then 
\begin{equation}
\label{ciclo}
\Sigma_0=\xi^i\frac{\partial {\cal L}}{\partial {\dot q}^i}\,,
\end{equation}
as a consequence of $L_X{\cal L}=0$.
It should be emphasized that in this approach one starts considering the existence of an arbitrary Noether symmetry (\ref{X}), and the particular form of this symmetry is obtained once one imposes $L_X{\cal L}=0$ for a particular class of Lagrangians.
Furthermore,  the energy function
\begin{equation}\label{energy}
 E_{\cal L}(q_i,\,\dot q_i)=\frac{\partial {\cal L}}{\partial \dot q^i}\dot q^i-{\cal L}\,,
\end{equation}
is conserved as well. The right hand side of equation (\ref{energy}) corresponds to the Hamiltonian of the system, $\cal H$, and given that $\partial \cal{H}/\partial$$t=$-$\partial \cal{L}/\partial$$t$, 
it follows immediately that $E_{\cal L}$ is conserved, whenever $\cal{L}$ does not depend on time. Therefore, the use of this equation further simplifies the problem and analytic solutions can be easily found.
This approach has been used several times in cosmology giving  exact solutions of physical interest (see for example \cite{felicia,basilakos,defelice,nesseris}). Although in most cases the particular symmetry has not a clear physical interpretation, it can select a model from the corresponding family of theories.

%%%%%%%%%%%%%%%%%%%%%%%%%%%%%%%%%
%%%%%%%%%%%%%%%%%%%%%%%%%%%%%%%%%
\subsection{Noether symmetries for bigravity cosmology}\label{s32}

Now that we have a point-like Lagrangian that is non-degenerate, we can look for the potential existence of  Noether symmetries.
Taking into account equation (\ref{X0}), a general vector field on ${\cal TQ}\equiv\{a,\dot{a},b,\dot{b}\}$ can be expressed as
\begin{equation}\label{X}
 X=\xi(a,b)\frac{\partial}{\partial a}+\eta(a,b)\frac{\partial}{\partial b}+
\frac{d\xi(a,b)}{dt}\frac{\partial}{\partial \dot a}+
\frac{d\eta(a,b)}{dt}\frac{\partial}{\partial \dot b}.
\end{equation}
The point-like Lagrangian (\ref{pl2}) has a Noether symmetry if the condition (\ref{Lie0}) is satisfied for the vector field (\ref{X}). The Noether symmetry is at this point arbitrary, it will be fixed when $\xi(a,b)$ and $\eta(a,b)$ will be specified.

It can be shown that\footnote{We refer to the appendix \ref{appendixB} for the mathematical details.} the two components of the vector $X$ reads
\begin{eqnarray}\label{cambio}
 \xi(a,b)&=&\xi_0 a+\phi,\nonumber\\
\eta(a,b)&=&-\frac{3}{2\,\kappa}\xi_0\frac{a^2}{b}
-\frac{\phi}{\kappa}\frac{a}{b}+\frac{\delta}{b}-\frac{\xi_0}{2}b,
\end{eqnarray}
where $\xi_0$, $\phi$, $\delta$ are constants. The functions $\xi(a,b)$ and $\eta(a,b)$ must as well fullfill the conditions (\ref{cuatro}) and (\ref{seis}) given in the Appendix \ref{appendixB}. Those two equations will fix the matter content $B(b)=b^3\bar\rho_m(b)$ for a given matter content $A(a)=a^3\rho_m(a)$.

%%%%%%%%%%%%%%%%%%%%%%%%%%%%%%%%%
%%%%%%%%%%%%%%%%%%%%%%%%%%%%%%%%%
\subsection{Solutions compatible with matter in our universe}\label{s33}

Let us now take into account that in our universe there is only matter (dust) and radiation. That is
\begin{equation}\label{matter}
 A(a)=A_0+\frac{A_1}{a}.
\end{equation}
Therefore, we want to find which material content should be present in the other universe, $B(b)$,
to have a Noether symmetry for the biuniversal dynamics, given by equations (\ref{X}) and (\ref{cambio}) (cf. Appendix \ref{appendixB}).
Hence, we look for functions $B(b)$ that are solutions of both equations (\ref{cuatro}) and (\ref{seis}),
once expressions (\ref{matter}) and (\ref{cambio}) have been introduced in these equations.
The solutions that we present in the following discussion also restrict some of the $\beta_i$ parameters and the function $A(a)$ given by equation (\ref{matter}). Restriction of the $\beta_i$ parameters means that only some of the ghost-free theories will be able to have a Noether symmetry for cosmological solutions, whereas further restriction of $A(a)$ implies that not all the matter and radiation contents are compatible with the existence of this symmetry. In particular, as we will show, compatibility of the system of differential equations requires $A_1=0$; therefore, a universe with a non-vanishing radiation contribution cannot have a Noether symmetry. However, as we will discuss in detail, a universe like ours that contains radiation can tend to a state with a Noether symmetry when the radiation component is sufficiently diluted. As we will show, this final state is of particular phenomenological interest.

%%%%%%%%%%%%%%%%%%%%%%%%%%%
\subsubsection{  General Relativity  with a cosmological constant}
The first solution is given by
\begin{equation}\label{parameters1}
 \xi_0=0,\qquad A_1=0,\qquad \beta_0=\beta_1=\beta_2=\beta_3=0.
\end{equation}
So the only non-vanishing parameter in the interaction Lagrangian is $\beta_4$, implying that
interaction between both sectors reduces to a cosmological constant in the second universe.  That is
\begin{equation}
 \rho_{\rm bg}(a/b)=0,\qquad\widebar\rho_{\rm bg}(b/a)=m^2\beta_4.
\end{equation}
In this case,
the dynamics of both universes is decoupled and have the same spatial geometry, i.e. the same value of $k$ (which is arbitrary).
The second universe is empty, since equations (\ref{cuatro}) and (\ref{seis}) (cf. Appendix \ref{appendixB}) imply that
the cosmological constant appearing in the geometric term cancel the contribution coming form $B(b)$ as
\begin{equation}\label{m1}
 B(b)=-b^3m^2\beta_4\Rightarrow\widebar\rho_{\rm m}(b)=-m^2\beta_4.
\end{equation}

Moreover, one can substitute the parameters in equations (\ref{parameters1}) and (\ref{m1}) into
equation (\ref{polinomio}) to study the solution of this system. This leads to
\begin{equation}
 \rho_{\rm m}(a) \,\frac{b}{a}=0.
\end{equation}
As $\rho_{\rm m}=A_0/a^3$, this implies that $b=0$ if $A_0\neq0$.
Assuming that this condition is preserved in time, one has $\dot{b}=0$ that implies
$\widebar N=0$ and, therefore, there is no second gravitational sector.
In this case, there is only one gravitational sector filled with dust matter and, therefore, there is no bigravity solutions for this model.
If one had $A_0=0$ instead of $b=0$, then the cosmological model with a Noether symmetry would correspond to a general relativistic model that is empty. 
Our Universe could tend to that state when the matter is infinitely diluted by the expansion.

%%%%%%%%%%%%%%%%%%%%%%%%%%%
\subsubsection{Putative bigravity solution}
The second solution for the system (\ref{cuatro}) and (\ref{seis}) (please cf. Appendix \ref{appendixB}) is given by
\begin{equation}
 \xi_0=0,\, A_1=0,\, \beta_1=\beta_3=0,\, \beta_0=\frac{3}{\kappa}\beta_2,\,\delta=0.
\end{equation}
In this case,  the interaction between both gravitational sectors is not equivalent to a cosmological constant,
since $\beta_2\neq0$. 
Indeed one has a non-trivial interaction affecting the dynamics of our universe, which is given by
\begin{equation}\label{rhobg1}
 \rho_{\rm bg}=3\beta_2m^2\left(\frac{1}{\kappa}+\frac{b^2}{a^2}\right).
\end{equation}
That is, a cosmological constant component plus a purely bimetric term.
It must be noted that the solution requires $A_1=0$, that is, there is no radiation in this universe. 
Our Universe contains, of course, relativistic matter, so this result could seem incompatible with the existence of a Noether symmetry in physical interesting situations. Nevertheless, it can be noted that the state with a Noether symmetry can be approached when the radiation component is diluted. So the universe could be in a state with a Noether symmetry in its asymptotic past or future; i.e. during  a ``pre-inflationary'' phase to be followed by a standard inflationary era or at the very late-time dark energy era. In both situations, the radiation contribution can be negligible as compared with the material content. Moreover, this solution is a solution valid for any $k$.

In the other universe, the material content compatible with the existence of a Noether symmetry is
\begin{equation}\label{cc2}
 \widebar\rho_{\rm m}=m^2\left(3\kappa\beta_2-\beta_4\right).
\end{equation}
Therefore, the other universe only contains vacuum energy when it is in the state with a Noether symmetry\footnote{One could be surprised about this non-symmetric result for the matter content given the symmetry of the original interaction term. Nevertheless, in order to obtain a non-degenerate and nontrivial Lagrangian we have shown that the
original symmetry should be broken in order to obtain a dynamical evolution.}, this corresponds to $\epsilon\neq0$ in action (\ref{Sintro}) but $\widebar\rho_{\rm m}$ leading to a cosmological constant contribution.
Hence, the solution for this particular model
corresponds to the ``minimal model'' discussed in \cite{vonStrauss:2011mq} plus an explicitly cosmological
constant contribution. This model was called minimal in reference~\cite{vonStrauss:2011mq} 
because, as $\beta_1=\beta_3=0$,  the
only nonlinear interaction term in both universes is the quadratic term (it should not be confused
with the minimal model introduced in reference~\cite{Konnig:2013gxa}). In this case one has that some
coefficients appearing in equation (\ref{polinomio}) are fixed, $c_0=c_2=c_4=0$. Furthermore, in our model  $\beta_0$ and $\beta_2$ are not independent and, therefore, one additionally has $c_1=0$. 

It is particularly simple to obtain the analytic solution using equation (\ref{polinomio}) and we do not need to use the Noether symmetry to find out the exact  solution, as in other alternative theories of gravity. \cite{Noether,Noether1,Noether2}.
Substituting these values of the parameters and equation (\ref{cc2}) in equation (\ref{polinomio}),
it can be seen that the solutions for this model are given by
\begin{equation}
 \rho_{\rm m}\,\frac{b}{a}=0.
\end{equation}
As in the previous case, if we take  $\rho_{\rm m}=A_0/a^3\neq0$, then we will have $b=0$. Thus, 
assuming $\dot{b}=0$, one can conclude that this model is not compatible with having a second gravitational sector. 
On the other hand, if one insists in considering $b\neq0$ to have a bigravity model, then our Universe will restore a state with a Noether symmetry when the matter component is infinitely diluted, that is $A_0=0$. In this case, however, $b(t)$ is left undetermined by the Friedmann equations; so $a(t)$ will depend on an undetermined function and we will not be able to conclude anything about the dynamics of our Universe (at least without assuming a form for $b(t)$).
Therefore, requiring the existence of a Noether symmetry leads either to a general relativistic world with a cosmological constant term
(fixed by $\beta_2$), which is suitable to describe the current acceleration of our universe, or to an empty model whose dynamics cannot be determined.

%%%%%%%%%%%%%%%%%%%%%%%%%%%
\subsubsection{Bigravity solution I}\label{sbgI}
The last set of solutions of the system (\ref{cuatro}) and (\ref{seis}) (cf. Appendix \ref{appendixB}) is
\begin{equation}\label{parameters2}
 \phi=0,\, A_1=0,\, \beta_1=\beta_3=k=0,\, \beta_0=\frac{6}{\kappa}\beta_2,\, \delta=0,
\end{equation}
which leads to a non-trivial interaction given by
\begin{equation}\label{int2}
 \rho_{\rm bg}(b/a)=3\beta_2m^2\left(\frac{2}{\kappa}+\frac{b^2}{a^2}\right).
\end{equation}
The content of the other universe is expressed as
\begin{equation}\label{m2}
 \widebar\rho_{\rm m}=-m^2\beta_4,
\end{equation}
therefore, we have again $\epsilon\neq0$ in action (\ref{Sintro}) but $\widebar\rho_{\rm m}$ is equivalent to a cosmological constant contribution.
As in the previous case, given the simplicity of the model corresponding to the parameters (\ref{parameters2})
and (\ref{m2}), we can easily solve equation (\ref{polinomio}) even before calculating the Noether symmetry and conserved quantity. Taking into account (\ref{parameters2})
and (\ref{m2}) in equation (\ref{polinomio}),  we obtain
\begin{equation}\label{Gamma2}
 \left(\frac{b}{a}\right)^2=-\frac{\rho_{\rm m}(a)}{3m^2\beta_2}-\frac{1}{\kappa}.
\end{equation}
Hence, on the one hand, in order to have a real function $b$, one  needs to consider $\beta_2<0$ and/or $\kappa<0$.
On the other hand, considering equation (\ref{Gamma2}) in the Friedmann equation (\ref{Fried-g}), or equivalently
in equation (\ref{Fried-f2}), with the definitions (\ref{rho-g}) and (\ref{rho-f}), one obtains
\begin{equation}\label{Hg2}
 H_g^2=\frac{m^2\beta_2}{\kappa M_P^2}.
\end{equation}
This implies that $\kappa$ and $\beta_2$ should have the same sign. Therefore, both parameters should be negative.
From equations (\ref{Gamma2}) and (\ref{Hg2}) it can be seen that, in this case, both $a$ and $b$ can be well defined. 
(Some comments regarding the physics of the other universe are included in  section \ref{gravitation}.)
Defining $M_{P_2}^2=-\kappa M_{P}^2>0$ and integrating equation (\ref{Hg2}), 
we obtain
\begin{equation}\label{a2}
 a(t)=a_0\, {\rm exp}\left(m\sqrt{|\beta_2|/M_{P_2}^2}t\right),
\end{equation}
So, even if we have considered the presence of matter in this universe ($A_0\neq0$), the evolution given by equation (\ref{a2}) is exactly de Sitter. The scale factor (\ref{a2})
can be re-written as
\begin{equation}\label{a2b}
 a(t)=a_0\, {\rm exp}\left(\sqrt{\Lambda_{\rm eff}/3}\,t\right),
\qquad \Lambda_{\rm eff}=3m^2\frac{|\beta_2|}{M_{P_2}^2}.
\end{equation}
Therefore, the effective cosmological constant which appears in equation (\ref{a2b}),
is different from the cosmological constant induced by the interaction term (\ref{rho-g}), which is 
$\Lambda_{\rm g}=m^2\beta_0$. Indeed, one has
\begin{equation}\label{L}
 \Lambda_{\rm eff}=\frac{\Lambda_{\rm g}}{2\,M_P^2},
\end{equation}
where in the case that one considers an explicit cosmological constant  $\Lambda$ in the action,
the term $\Lambda_{\rm g}$ has to  be replaced by $\Lambda_{\rm g}-\Lambda$.

On the other hand, the components of the Noether vector field  (\ref{X}) assume the explicit forms 
\begin{equation}
 \xi(a)=\xi_0\,a,\qquad \eta(a,\,b)=\frac{3M_P^2}{2M_{P_2}^2}\xi_0\frac{a^2}{b}-\frac{\xi_0}{2}b\,.
\end{equation}
Following the considerations in reference \cite{Noether}, the Noether  vector field can be written in a simplified form as 
\begin{equation}
 \tilde{X}=\frac{\partial}{\partial v},
\label{unitaryX}
\end{equation}
by defining a new par of  variables $\{u,\, v\}$  such that
$v$ is cyclic; i.e. the Lagrangian can be recast in a form such that $\tilde{\mathcal L}(u,\,\dot u,\, \dot v)$. This change of variables is always possible since there is a Noether symmetry and the condition (\ref{ciclo}) is satisfied.   The  change of variables can be realized assuming a regular transformation  $u = u(a, b)$ and $v = v(a, b)$ where the Jacobian of the transformation is different from zero.  
Thus, we have
\begin{equation}
 \xi(a)\frac{\partial u}{\partial a}+\eta(a,\,b)\frac{\partial u}{\partial b}=0,\qquad \xi(a)\frac{\partial v}{\partial a}+\eta(a,\,b)\frac{\partial v}{\partial b}=1,
\end{equation}
and  we can obtain
\begin{eqnarray}\label{cambio2}
 u(a,\,b)=u_0\, a\left(b^2-\frac{M_P^2}{M_{P_2}^2}a^2\right),\quad
v(a,\,b)=\frac{1}{\xi_0}\ln\left(a\right)+v_0\, a\left(b^2-\frac{M_P^2}{M_{P_2}^2}a^2\right),
\end{eqnarray}
where $u_0$ and $v_0$ are integration constants. Reverting (\ref{cambio2}) and assuming, without loss of
generality,  $v_0=u_0=\xi_0=1$, we obtain 
\begin{eqnarray}
\label{change}
 a=\exp\left(v-u\right),\quad
b=\sqrt{u\exp\left(u-v\right)+\frac{M_P^2}{M_{P_2}^2}
\exp\left[2\left(v-u\right)\right]}.
\end{eqnarray}
The Lagrangian, in terms of these new variables, assume the form $\tilde{\mathcal L}(u,\,\dot u,\, \dot v)$ where 
 the new variable $v$ is cyclic. From equation (\ref{unitaryX}), the constant of motion can be expressed as
\begin{equation}
 \Sigma_0=\frac{\partial \tilde{\mathcal L}}{\partial \dot v}\,,
\end{equation}
which is the same as that defined in equation (\ref{ciclo}) but expressed in the new variables.
This quantity can be rewritten in terms of the scale factor of both universes as
\begin{eqnarray}\label{Sigma0}
 \Sigma_0&=&3\, M_{P_2}^2 a\,b\,\dot b-\frac{3}{2}\left[M_{P_2}^2 b^2+M_P^2 a^2\right]+3\beta_2m^2\frac{a^3\,b\,\dot b}{\dot a^2}\nonumber\\
&-& \frac{3\beta_2 m^2}{2}\frac{a^2}{\dot a}\left(3\frac{M_P^2}{M_{P_2}^2}a^2-b^2\right).
\end{eqnarray}
The fact that this quantity is conserved can be used to integrate the equations of motion. In our case,
however, they were easy to integrate so one could check that $\Sigma_0$ is indeed conserved using the solutions.
Taking into account the equations (\ref{a2}) and  (\ref{change}), equation (\ref{Sigma0}) simplifies to
\begin{equation}
 \Sigma_0=-\frac{A_0 \,M_{P_2}}{m\,|\beta_2|^{1/2}}=-\frac{A_0}{m}\sqrt{3/\Lambda_{\rm eff}}.
\end{equation}
Therefore, the conserved quantity associated to the Noether symmetry fixes the value of the effective cosmological constant $\Lambda_{\rm eff}$. The existence of a Noether symmetry for this solutions implies that the model evolves exactly as a de Sitter space even when matter has not been completely diluted.
It has to  be emphasized that, of course, our Universe contains both matter and radiation. However, we consider that it could approach the state with a Noether symmetry when the radiation component is completely negligible as compared with the matter component. 
In particular, it tends to a state where the effect of matter in the dynamics is negligible even before $\Omega_{\rm m}$ is small enough to assume that the material content is diluted. 

%%%%%%%%%%%%%%%%%%%%%%%%%%%
\subsubsection{Bigravity solution II}\label{sbgII}
The last set of solutions of the system (\ref{cuatro}) and (\ref{seis}) is given by
\begin{eqnarray}\label{parameters3}
 \phi=0,\, A_1=0,\, \beta_1=\beta_3=0,\,
 \beta_0=\frac{\beta_2}{\kappa},
\, k=\frac{2\beta_2m^2\delta}{M_P^2\,\xi_0}.
\end{eqnarray}
For this case, the interaction term in our universe has the same form as the one given in equation (\ref{int2}), that is
\begin{equation}
 \rho_{\rm bg}(b/a)=3\beta_2 m^2\left(\frac{2}{\kappa}+\frac{b^2}{a^2}\right),
\end{equation}
but now the material content of the other universe is
\begin{equation}
 \widebar\rho_{\rm m}(b)=-m^2\beta_4+\frac{6\kappa m^2\beta_2\delta}{\xi_0}b^{-2},
\end{equation}
which is equivalent to consider a cosmological constant contribution plus a spatial curvature contribution.
Therefore, from equation (\ref{polinomio}),  we get (considering  $A_0$  positive, see Eq.~(3.22))
\begin{equation}\label{b3}
 b=\sqrt{\frac{2\delta}{\xi_0}-\frac{A_0}{3m^2\beta_2}\frac{1}{a}-\frac{1}{\kappa}a^2},
\end{equation}
which can be introduced in the Friedmann equation (\ref{Fried-g}) to obtain
\begin{equation}\label{H3}
 H_g^2=\frac{\beta_2m^2}{\kappa M_P^2}.
\end{equation}
Assuming that the state with a Noether symmetry is approached at later times in our universe, that is for large
$a$, and taking into account equation (\ref{b3}) and (\ref{H3}) one  needs to consider negative values\footnote{In this case one could also have well-defined solutions for positive values of $\beta_2$ and $\kappa$ by restricting attention to small values of the scale factor $a$. We are interested, however, in the situation in which the solution having a Noether symmetry is approached at late-time. Notice that a positive value of $\beta$ would imply a minimum value for the scale factor which would lead to an avoidance of the Big Bang singularity in our universe. Nevertheless, as we said before, we are mainly interested on the late-time behavior of the universe.}
of $\beta_2$ and $\kappa$, and one can define again $M_{P_2}^2=-\kappa M_p^2$. As in the previous case,
we obtain
\begin{equation}\label{a3}
 a(t)=a_0\, {\rm exp}\left(\sqrt{\Lambda_{\rm eff}/3}\,t\right),
\qquad \Lambda_{\rm eff}=3m^2\frac{|\beta_2|}{M_{P_2}^2},
\end{equation}
although in this case this accelerating universe could have a non-vanishing spatial curvature as shown in equation
(\ref{parameters3}), its contribution to the dynamics is cancelled by the interaction term.

In this case,  the vector field associated with the Noether symmetry, is given by
\begin{equation}
\xi(a)=\xi_0 a,\qquad  \eta(a,\,b)=-\frac{\xi_0}{2}b+\frac{3M_P^2\xi_0}{2M_{P_2}^2}\frac{a^2}{b}+\frac{\delta}{b}.
\end{equation}
For the the cyclic  variable, one can again perform the transformation  $\{a,b\}\rightarrow\{u,v\}$ obtaining 
\begin{eqnarray}\label{cambio3}
 u(a,\,b)&=&u_0\, a\left(b^2-\frac{M_P^2}{M_{P_2}^2}a^2-\frac{2\delta}{\xi_0}\right),\\
v(a,\,b)&=&\frac{1}{\xi_0}\ln\left(a\right)+v_0\, a\left(b^2-\frac{M_P^2}{M_{P_2}^2}a^2-\frac{2\delta}{\xi_0}\right),
\end{eqnarray}
where $u_0$ and $v_0$ are integration constants.
Assuming $v_0=u_0=\xi_0=1$ in (\ref{cambio3}), one obtains
\begin{eqnarray}
 a=\exp\left(v-u\right),\quad
b=\sqrt{u\exp\left(u-v\right)+\frac{M_P^2}{M_{P_2}^2}
\exp\left[2\left(v-u\right)\right]+2\delta}.
\end{eqnarray}
Expressing the point-like Lagrangian in terms of $ u(a,\,b)$ and $v(a,\,b)$, it can be checked that $v$ is the
cyclic variable. As above,  there is a conserved quantity which, taking into account the solutions (\ref{b3})
and (\ref{a3}) (cf. Appendix \ref{appendixB}),  reduces to
\begin{equation}
 \Sigma_0=-\frac{A_0 \,M_{P_2}}{m\,|\beta_2|^{1/2}}=-\frac{A_0}{m}\sqrt{3/\Lambda_{\rm eff}}.
\end{equation}
As in the previous case the conserved quantity associated to the Noether symmetry fixes the value of the effective cosmological constant $\Lambda_{\rm eff}$. 

%%%%%%%%%%%%%%%%%%%%%%%%%%%
%%%%%%%%%%%%%%%%%%%%%%%%%%%
\section{The physics of the other universe}\label{gravitation}
Before discussing the physics of the second universe,  it should be emphasized that the biuniversal interpretation
of solutions of bigravity theory is mainly based on considering this theory as a fundamental representation of
reality. We consider this interpretation following the spirit of the weakly coupled worlds of reference \cite{Damour:2002ws}.
However, if one considers bigravity as an effective description of the physics of our universe at a given range of energies  instead, then it would not make any sense to refer to the physics of the other universe. In this section, we will investigate the the physics of the other universe as a physical realizable world. We will first discuss some potential issues of the bigravity models obtained in the previous section in subsection \ref{spi} and study the cosmological evolution of both models in subsections \ref{scI} and \ref{scII}.

%%%%%%%%%%%%%%%%%%%%%%%%%%%
%%%%%%%%%%%%%%%%%%%%%%%%%%%
\subsection{Potential issues}\label{spi}
The bigravity solutions with a Noether symmetry obtained in \ref{sbgI} and \ref{sbgII} require that the  gravitational coupling for each universe
is given by
\begin{equation}
 G_g=\frac{1}{8\pi M_P^2},\qquad G_f=-\frac{1}{8\pi M_{P_2}^2},
\end{equation}
respectively. Therefore, whereas the gravitational coupling is positive in our universe, it is negative in the other universe. So the stability of these models should be considered in further detail; in particular, the potential existence of a spin-2 ghost.

On the other hand, when the bi-universe has reached the state with a Noether symmetry studied in the previous section,
it would not be any material content in the second universe to feel this anti-gravitational field. Nevertheless, we consider that some material content was present in earlier cosmological phases,
but it has been diluted by the cosmic expansion. This matter would follow geodesics of the metric $f_{\mu\nu}$, since
the matter fields are minimally coupled to this metric. Nevertheless, the way in which the material 
content (together with the coupling between both gravitational sectors) would 
curve the geometry is ``the contrary'' than in our universe, that is
\begin{eqnarray}
 G^{\mu}{}_{\nu} &=&8\pi G_g \left(T^{({\rm m})\mu}{}_{\nu}+T^{\mu}{}_{\nu}\right),\\
\widebar{G}^{\mu}{}_{\nu} & =&
-8\pi |G_f|\left(\widebar T^{({\rm m})\mu}{}_{\nu} + \widebar{T}^{\mu}{}_{\nu}\right).
\end{eqnarray}
Therefore, it seems that in the second universe there would be a repulsive gravitational field, but all particles would fall
the same way in this field in agreement with the equivalence principle. It is worth noticing that   the $(-)$ sign in front of $\widebar T^{({\rm m})\mu}{}_{\nu}$ 
can be eliminated by explicitly re-introducing the parameter $\epsilon$ of equation (\ref{Sintro}) on the last term of the action (\ref{action}), where $\epsilon=-1$.  In such a way, the symmetry between the two universes is fully restored.

%%%%%%%%%%%%%%%%%%%%%%%%%%%
%%%%%%%%%%%%%%%%%%%%%%%%%%%
\subsection{Cosmological evolution of the first bigravity model}\label{scI}
The cosmological evolution of the second universe described by the first solution, solution \ref{sbgI}, can be obtained by taking into account
equations (\ref{Gamma2}) and (\ref{a2}). This is
\begin{equation}\label{b2}
 b(t)=b_0\, {\rm exp}\left(\alpha\,t\right)
\sqrt{1+\gamma\,{\rm exp}\left(-3\alpha\,t\right)},
\end{equation}
where
\begin{equation}
 b_0=a_0\frac{M_P}{M_{P_2}},\, \alpha=\frac{m\,|\beta_2|^{1/2}}{M_{P_2}},\,
\gamma=\frac{A_0}{3M_P^2\alpha^2a_0^3}.
\end{equation}
The scale factor (\ref{b2}) can be written in terms of the cosmic time of the second universe, $\tau$, to be suitably interpreted.
This cosmic time is given by
\begin{equation}\label{taui}
 \tau=\int \widebar N dt=\int \frac{\dot b}{\dot a} dt.
\end{equation}
Deriving equations (\ref{a2}) and (\ref{a2b}), and then integrating (\ref{taui}), this cosmic time is
\begin{eqnarray}\label{tau2}
 \tau=\tau_0\left\{3\alpha t+\sqrt{1+\gamma \,e^{-3\alpha t}}
+ 2\ln\left(1+\sqrt{1+\gamma \,e^{-3\alpha t}}\right)+C\right\},
\end{eqnarray}
where 
\begin{equation}
 \tau_0=\frac{b_0 M_{P_2}}{3 a_0 m\,|\beta_2|^{1/2}},
\end{equation}
and $C$ is an integration constant which can be chosen to impose some origin between both times (as $\tau(0)=0$).
Although a complete analytic expression of $b(\tau)$ cannot be easily obtained from equations (\ref{b2})
and (\ref{tau2}), for large
$t$ one has
\begin{equation}
 b\approx 
 b_0 \exp\left[\sqrt{\frac{\widebar\Lambda_{\rm eff}}{3}}\left(\tau-\tau_*\right)\right],
\end{equation}
where
\begin{equation}
 \widebar\Lambda_{\rm eff}=\frac{3\,m^2\,|\beta_2|}{M_P^2},
\end{equation}
Taking into account equation (\ref{a2b}), this effective cosmological constant can be written as
\begin{equation}
 \widebar\Lambda_{\rm eff}=\frac{M_{P_2}^2}{M_P^2}\Lambda_{\rm eff},
\end{equation}
and
\begin{equation}
\tau_*=\frac{1}{\sqrt{3\widebar\Lambda_{\rm eff}}}\left[\sqrt{1+\gamma}+2\ln\left(1+\sqrt{1+\gamma}\right)+C\right].
\end{equation}
Therefore, although the evolution of the second universe is not exactly de Sitter, it will approach a de Sitter behaviour asymptotically. 

%%%%%%%%%%%%%%%%%%%%%%%%%%%
%%%%%%%%%%%%%%%%%%%%%%%%%%%
\subsection{Cosmological evolution of the second bigravity model}\label{scII}
For the second solution presented in \ref{sbgII}, taking into account equations (\ref{b3}) and (\ref{a3}) (cf. Appendix \ref{appendixB}), one has
\begin{equation}\label{b3t}
 b(t)=\sqrt{b_1\,{\rm exp}(2\alpha t)+b_2\,{\rm exp}(-\alpha t)+2\xi_0},
\end{equation}
with 
\begin{equation}
 b_1=a_0^2\frac{M_P^2}{M_{P_2}^2},\,b_2=\frac{A_0}{3|\beta_2|m^2a_0},\,\alpha=\frac{m|\beta_2|^{1/2}}{M_{P_2}}.
\end{equation}
Although we have not being able to obtain an analytic expression for $\tau(t)$ in this case, it can be noted
that the first term in equation (\ref{b3t}) dominates for large $t$. Therefore, the late time behaviour is similar
to that obtained in the previous section; i.e. the second universe will be again asymptotically de Sitter in the future.

%%%%%%%%%%%%%%%%%%%%%%%%%%%
%%%%%%%%%%%%%%%%%%%%%%%%%%%
\section{Discussion and Conclusions}\label{summary}
Bigravity theories are those theories of gravitation where two dynamical and mutually interacting metrics are considered. 
They represent a natural approach to avoid background dependence and ghosts in massive gravity and in other fundamental theories of gravitational interaction.
In this framework, when two sets of matter fields minimally coupled to each metric are considered, one can interpret this scenario as two classical universes that are interacting only through their gravitational sectors if bigravity is considered a fundamental description of reality. Anyway, as the matter content of the second sector cannot be measured in a direct way, the dynamics of our Universe seems to be determined by some unobservable (and, therefore, arbitrary) quantity.

In this paper we have assumed that the material content minimally coupled to the second sector is not arbitrary, but it has to be such that the bigravity cosmology has a Noether symmetry. Under this assumption we have obtained four particular cosmological models. 
In the first model the two sectors decoupled and the theory describing our universe is equivalent to GR with a cosmological constant.
Although the second model seems a genuine bigravity model at first sight, we have shown that it also reduces to   GR when we take into account that our Universe is not empty.
Therefore, these two models point out to the fact that we have an additional symmetry in   GR as compared to bigravity, at least in cosmological scenarios.
In the third place, we have also obtained two genuine bigravity models with a Noether symmetry, both leading to an exact de Sitter evolution for our Universe even if there is dust matter minimally coupled to our gravitational sector. The conserved quantity associated to the Noether symmetry provides us with the effective cosmological constant of this de Sitter evolution. The Noether symmetry, therefore, enforces a de Sitter evolution, which is of special phenomenological interest to describe the late-time accelerated cosmological expansion. It should be noted that we are not stating that all de Sitter solutions are compatible with a Noether symmetry, but only the particular solutions presented in the paper for the very particular bigravity models obtained.

Regarding the genuine bigravity models some relevant issues should be emphasized. 
We have obtained that the existence of non-vanishing radiation minimally coupled to our gravitational sector is incompatible with a Noether symmetry. This fact may be taken against the consideration of Noether symmetries as a fundamental criterium to select the cosmological evolution, as our real universe contents radiation. Nevertheless, we interpret the absence of radiation in the Noether symmetric solutions as indicating that the Noether symmetric state is approached with time. That is, our universe will approach a de Sitter state with a Noether symmetry when the radiation component is infinitely diluted. 
On the other hand, the gravitational couplings of both sectors have different signs. This different sign, which may lead to potential instabilities, could imply the existence of anti-gravitational effects in the other sector, 
when it is interpreted as a second weakly coupled universe. 
{If those instabilities are confirmed by a more careful analysis, the existence of a Noether symmetry would signal the need of decoupling
both gravitational sectors and, therefore, support GR as the most consistent theory of gravity at a fundamental level.}

Taking into account the bigravity cosmological models already studied in the literature, it can be noted that both our bigravity cosmologies require $\beta_1=\beta_3=0$. This model has been studied in detail in reference \cite{vonStrauss:2011mq} assuming an empty second gravitational sector. It was shown that the model was equivalent to GR with a rescaled Planck mass appearing in the modified Friedmann equation multiplying the energy density and a shifted cosmological constant. However, in our bigravity models we consider the existence of matter in the second gravitational sector. This matter content just produces additional constant terms in the numerator and denominator of the rescaled Planck mass when it has a constant energy density (with an additional contribution proportional to $k/b^2$ if $k\neq0$). The value of that constant terms are fixed in our particular models in such a way that the rescaling factor vanishes. Hence, the matter minimally coupled to our sector does not contribute to the dynamics of our universe and so we have obtained an exact de Sitter evolution for both models even if there is matter present. The states with enhanced symmetry are, therefore, not only equivalent to a GR solution, but they are just a de Sitter space with the rescaled vacuum energy given by the conserved quantity.

Furthermore, it can be noted that our solutions have some resemblance to the partially massless model \cite{Hassan:2012gz}, since $\beta_1=\beta_3=0$ and $ \beta_0\propto\beta_2$. However, they do not coincide because, on one hand, in our case $\beta_4$ is not fixed and, on the other hand, we restrict the possible matter content. Hence, our models are more general and more restrictive at the same time. However, it is interesting to note the overlapping in the parameter space of both models which have enhanced symmetries.

It should be noted that our application of the Noether Symmetry Approach to bigravity is different of the use of this approach in other alternative theories of gravity. Here we have adopted another strategy. Instead of determining the form of the point-like Lagrangian and then searching for exact solutions by the Noether Symmetry Approach, we have determined the matter content and the evolution of the two interacting universes asking for the existence of the Noether symmetry.
A further interesting   point about Noether symmetry in bigravity  models is that whenever there are two  universes the model is free from singularities. In other words, the presence of symmetries seems to remove singularities.

The role of Noether symmetries in the quantum framework has been discussed in reference \cite{Noether2}. Considering the wave function of the universe in quantum cosmology, the Hartle criterion suggests that peaks of the wave function of the universe imply the existence of correlations among the geometrical and matter degrees of freedom and, therefore, the emergence of classical trajectories (i.e. universes). As it has been proven in \cite{Noether2}, for some theories of gravity the wave function is picked along the trajectories with Noether symmetries, the absence of Noether symmetries gives rise to non-correlated variables and then to non-observable universes. 

In future projects, we will consider the stability of the obtained bigravity models as well as the possible observational signatures of consistent models that approach the cosmological state with a Noether symmetry. Moreover, we will also investigate the quantum cosmology of the present model following, for example, the approach used in reference \cite{Bouhmadi-Lopez:2016dcf} within a Palatini type of theory that can be regarded as a bimetric scenario.

%%%%%%%%%%%%%%%%%%%%%%%%%%%%%%%%%%%%%%%%%%%%%%%%%%%%%
\begin{acknowledgements}
The work of MBL is supported by the Basque Foundation of Science Ikerbasque. She also wishes to acknowledge the partial support from the Basque government Grant No. IT956-16 (Spain) and FONDOS FEDER under grant FIS2014-57956-P (Spanish government). This research work is supported partially by the Portuguese grand UID/MAT/00212/2013. 
SC acknowledges INFN Sez. di Napoli (Iniziative Specifiche QGSKY and TEONGRAV).  
PMM acknowledges financial support from the Spanish Ministry of Economy and Competitiveness through the postdoctoral training contract FPDI-2013-16161 and the project FIS2014-52837-P, and from AEI (Spain) and FEDER under the project FIS2016-78859-P(AEI/FEDER, UE).
This article is based upon work from COST Action (CA15117, CANTATA), supported by COST (European Cooperation in Science and Technology).
\end{acknowledgements}
%%%%%%%%%%%%%%%%%%%%%%%%%%%%%%%%%%%%%%%%%%%%%%%%%%

\appendix
\section{Symmetric gauge fixing}\label{symmetric}
When looking for a non-degenerate Lagrangian, we could have chosen to fix the gauge in Lagrangian (\ref{pl1}) without breaking the symmetry under inter-change
of both sectors. The simplest gauge fixing in agreement with equation (\ref{Bi}) is
\begin{equation}\label{a1}
 N=\dot{a},\qquad {\rm and}\qquad \widebar{N}=\dot{b}.
\end{equation}
In this case the point-like Lagrangian is
\begin{eqnarray}\label{pl3}
 {\cal L}&=&- \left[3M_P^2\,a\,\dot a\,(1-k)+a^3\dot{a}\rho_{\rm bg}(b/a)+a^3\dot{a}\rho_{\rm m}(a)\right]\nonumber \\
&-&\left[3M_f^2\,b\,\dot b\,(1-k)+b^3\,\dot b\widebar\rho_{\rm bg}(a/b)+b^3\dot b\widebar\rho_{\rm m}(b)\right] ,
\end{eqnarray}
{where all  terms are linear either in $\dot{a}$ or in $\dot{b}$. Due to this linearity,  the determinant of the Hessian of this Lagrangian
vanishes and, therefore, it is degenerate.}

{If one insists in studying the dynamics of this degenerate Lagrangian with non redundant degrees of freedom,}
the equations of motion of this Lagrangian can be obtained by considering the variation with respect to $a$ and
the variation with respect to $b$. It can be seen that both cases lead to
\begin{equation}\label{a3}
 \widebar\rho_{\rm bg}'(a/b)=\frac{a^3}{b^3}\rho_{\rm bg}^s(b/a),
\end{equation}
which is not a dynamical equation. Moreover, equation (\ref{a3}) is simply an identity satisfied by the effective energy densities, which are given by (\ref{rho-g}) and (\ref{rho-f}).
On the other hand, as all the terms in the Lagrangian (\ref{pl3}) are linear either in $\dot{a}$ or in $\dot{b}$, 
the energy function (\ref{energy}) is exactly zero and, therefore, the constraint provided by requiring a constant energy is also trivially satisfied.

In summary, by fixing a gauge leading to a symmetric point-like Lagrangian, we have obtained a trivial Lagrangian which does not generate any
dynamical equation. It should be noted that the physical properties of the system cannot change, of course, by a gauge fixing. The Lagrangian 
(\ref{pl3}) does not imply physical characteristics different from those described by The Lagrangian  (\ref{pl1}), it simply does not provide us 
with any information {about the system because it is degenerate even if we have removed the redundant variables of the phase space}. 
Hence, the choice (\ref{a1}) is a bad gauge choice that renders the Lagrangian trivial.

\section{Noether symmetries for bigravity in FLRW Universes: general matter content}\label{appendixB}

We next explain how the the vector field $X$ defined in Eq.~(\ref{X}) that define the Noether Symmetry can de obtained. First of all by applying the condition (\ref{Lie0}) to the point-like Lagrangian (\ref{pl2}), one obtains an expression depending on $\xi(a,b)$ and $\eta(a,b)$ and their derivatives that has to vanish, that is $ L_X{\cal L}(a,\, b,\,\dot a,\,\dot b)=0$.
Imposing that the coefficients of the terms  $\dot a^2$, $\dot b^2$, $\dot a \dot b$, $\dot b/\dot a$, $\dot b^2/\dot a^2$
and $\dot b^0/\dot a^0$ appearing in this expression all vanish, we obtain a system of differential equations for $\xi(a,b)$ and $\eta(a,b)$  (see reference \cite{Noether} for details). These are:
\begin{equation}\label{uno}
 \xi+2\,a\,\xi'+\kappa\,b\,\eta'=0,
\end{equation}
\begin{equation}\label{dos}
 \xi^s=0,
\end{equation}
\begin{equation}\label{tres}
 \eta+\frac{2}{\kappa}a\,\xi^s+b\xi'+b\,\eta^s=0,
\end{equation}
\begin{eqnarray}\label{cuatro}
&&\eta\,B^s+(\eta^s-\xi')B +\widebar C'\xi +\widebar C^s\eta+\widebar C(\eta^s-\xi')\nonumber\\
&+&3\kappa M_P^2\,k(-\eta+b\,\xi'-b\,\eta^s)=0,
\end{eqnarray}
\begin{equation}\label{cinco}
 (-3\,\kappa M_P^2b\,k+\widebar C+B)\xi^s=0,
\end{equation}
and
\begin{equation}\label{seis}
 A'\xi+B\eta'+C^s\eta+C'\xi+\widebar C\eta'-3\,M_P^2\,k\xi-3\kappa M_P^2\,k\,b\eta'=0,
\end{equation}
where we have defined  $'\equiv\partial/\partial a$, ${}^s\equiv\partial/\partial b$, $C(a,b)=a^3\rho_{\rm bg}(b/a)$,
$\widebar C(a,b)=b^3\widebar\rho_{\rm bg}(a/b)$, $A(a)=a^3\rho_{\rm m}(a)$ and $B(b)=b^3\widebar\rho_{\rm m}(b)$.
It should be emphasized that we have not introduced any particular form for the functions $\rho_{\rm m}(a)$ and $\widebar \rho_{\rm m}(b)$ in the Lagrangian (\ref{pl2}); we just consider that they are given by the continuity equation once a particular kind of material content is assumed. 
We can now wonder if there are some functions $\rho_{\rm m}(a)$ and $\widebar \rho_{\rm m}(b)$ for which the system is compatible;
in other words, we study if some material contents minimally coupled to each gravitational sectors are compatible with the existence of a Noether symmetry.
In summary, the unknown functions\footnote{It is worth noticing that if the configuration space $\cal Q$ has dimension $n$, the number of differential equations emerging from $L_X{\cal L}=0$ is $1+n(n+1)/2$. In our case, we obtain six equations instead of four, because we have two more unknown functions that are $\rho_{\rm m}(a)$ and $\widebar \rho_{\rm m}(b)$. } of the system of equations (\ref{uno})-(\ref{seis})
are $\xi(a,b)$, $\eta(a,b)$, $\rho_{\rm m}(a)$ and $\widebar \rho_{\rm m}(b)$. Hence, the system has more equations than unknown functions and can be, therefore, incompatible. If that was the case and there were no solution, then one would conclude that there is no minimally coupled material content that allows the existence of a Noether symmetry in bigravity cosmology (at least, there would not be a Noether symmetry assuming Einstein-Hilbert gravitational terms and the fulfillment of the equivalence principle).

Equation (\ref{dos}) simply implies that
\begin{equation}\label{dosb}
 \xi^s=0\Rightarrow\xi=\xi(a).
\end{equation}
This condition makes equation (\ref{cinco}) trivially satisfied.
Taking into account that $\xi$ is independent of $b$ in equation (\ref{uno}), one has
\begin{equation}\label{unob}
\eta'(a,\,b)=\frac{\eta_0'(a)}{b}\Rightarrow\eta(a,\,b)=\frac{\eta_0(a)}{b}+c(b),
\end{equation}
being $\eta_0(a)$ a function of $a$ which should be determined using the other equations.
Introducing equations (\ref{unob}) and (\ref{dosb}) in equation (\ref{tres}), one gets
\begin{equation}
 \frac{c(b)}{b}+c^s(b)=-\xi'(a)\equiv -\xi_0,
\end{equation}
which must, therefore, be equal to a constant that we have denoted as $-\xi_0$. Then, this implies
\begin{equation}
 \xi(a)=\xi_0 a+\phi,\qquad \eta(a,b)=\frac{\eta_0(a)}{b}-\frac{\xi_0}{2}b+\frac{\delta}{b},
\end{equation}
where $\phi$ and $\delta$ are arbitrary constants. Now, considering again equation (\ref{uno}), we have
\begin{eqnarray}\label{cambio}
 \xi(a)&=&\xi_0 a+\phi,\nonumber\\
\eta(a,b)&=&-\frac{3}{2\,\kappa}\xi_0\frac{a^2}{b}
-\frac{\phi}{\kappa}\frac{a}{b}+\frac{\delta}{b}-\frac{\xi_0}{2}b.
\end{eqnarray}
Therefore, the most general $\xi(a)$ and $\eta(a,\,b)$ compatible with equations (\ref{uno}), (\ref{dos}),
(\ref{tres}) and (\ref{cinco}) are given by (\ref{cambio}). In order to obtain these functions,  we have not
imposed any restrictions on $A(a)$ and $B(b)$, but we have still two equations of the system.
Hence, the system will be compatible if there are functions $A(a)$ and $B(b)$ which are solutions of
equations (\ref{cuatro}) and (\ref{seis}) with $\xi(a)$ and $\eta(a,\,b)$ given by equation (\ref{cambio}).

%%%%%%%%%%%%%%%%%%%%%%%%%%%%%%%%%%%%%%%%%%%%%%%%%%%%%%

\end{document}